\documentclass[twoside,leqno,twocolumn]{article}

\usepackage[letterpaper]{geometry}
\usepackage{graphicx}
\usepackage{ltexpprt}
\usepackage{hyperref}
\usepackage{subcaption}
\usepackage{bm}
\usepackage{amsmath}
\usepackage{adjustbox}
\usepackage{multirow}
\usepackage{color}

\begin{document}

\newcommand\relatedversion{}

\title{\Large Localized Graph Collaborative Filtering}
\author{Yiqi Wang\thanks{Michigan State University. {wangy206,jinwei2,tangjili}@msu.edu. }
\and Chaozhuo Li\thanks{ Microsoft Research Asia . {cli,xingx}@microsoft.com.}
\and Mingzheng Li\thanks{ Microsoft. {mingzhengli,yumliu,hasun}@microsoft.com.}
\and Wei Jin\footnotemark[1] 
\and Yuming Liu\footnotemark[3] 
\and Hao Sun\footnotemark[3] 
\and Xing Xie\footnotemark[2] 
\and Jiliang Tang\footnotemark[1] 

}
\date{}

\maketitle


\fancyfoot[R]{\scriptsize{Copyright \textcopyright\ 2022 by SIAM\\
Unauthorized reproduction of this article is prohibited}}





\begin{abstract} \small\baselineskip=9pt User-item interactions in recommendations can be naturally denoted as a user-item bipartite graph. Given the success of graph neural networks (GNNs) in graph representation learning, GNN-based Collaborative Filtering (CF) methods have been proposed to advance recommender systems. These methods often make recommendations based on the learned user and item embeddings. However, we found that they do not perform well with sparse user-item graphs which are quite common in real-world recommendations. Therefore, in this work, we introduce a novel perspective to build GNN-based CF methods for recommendations which leads to the proposed framework Localized Graph Collaborative Filtering (LGCF). One key advantage of LGCF is that it does not need to learn embeddings for each user and item, which is challenging in sparse scenarios. Alternatively, LGCF aims at encoding useful CF information into a localized graph and making recommendations based on such graph. Extensive experiments on various datasets validate the effectiveness of LGCF, especially in sparse scenarios. Furthermore, empirical results demonstrate that LGCF provides complementary information to the embedding-based CF model which can be utilized to boost recommendation performance. \end{abstract}

\section{Introduction.}\label{sec:introduction}
The rapid development of information technology has facilitated an explosion of information, and it has accentuated the challenge of information overload. 
Recommender systems aim to mitigate the information overload problem by suggesting a small set of items for users to meet their personalized interests. They have been widely adopted by various online services such as e-commerce and social media~\cite{okura2017embedding}. 
The key to build a personalized recommender system lies in modeling users’ preference on items based on their historical interactions (e.g., ratings and clicks), known as collaborative filtering (CF)~\cite{su2009survey}.  
In recommendation systems, the user-item interactions can naturally form a bipartite graph and GNNs have shown great potential to further improve the CF performance~\cite{wang2019neural,he2020lightgcn}. GNNs are capable of capturing the CF signals encoded in the graph topology. Specifically, high-order user-item interactions can be explicitly captured by the stacked GNN layers, which contributes to learning expressive representations.
For example, Neural Graph Collaborative Filtering (NGCF)~\cite{wang2019neural} stacks several embedding propagation layers to explicitly encode the high-order connectivity in the user-item graph into the embedding process. LightGCN~\cite{he2020lightgcn} simplifies the design for GNNs for recommendation by only including neighborhood aggregation while eliminating other operations such as feature transformation.

\begin{figure}[t]
  \centering
  \subfloat[Tianchi-2798696\label{fig:intro-27}]{{\includegraphics[width=0.5\linewidth]{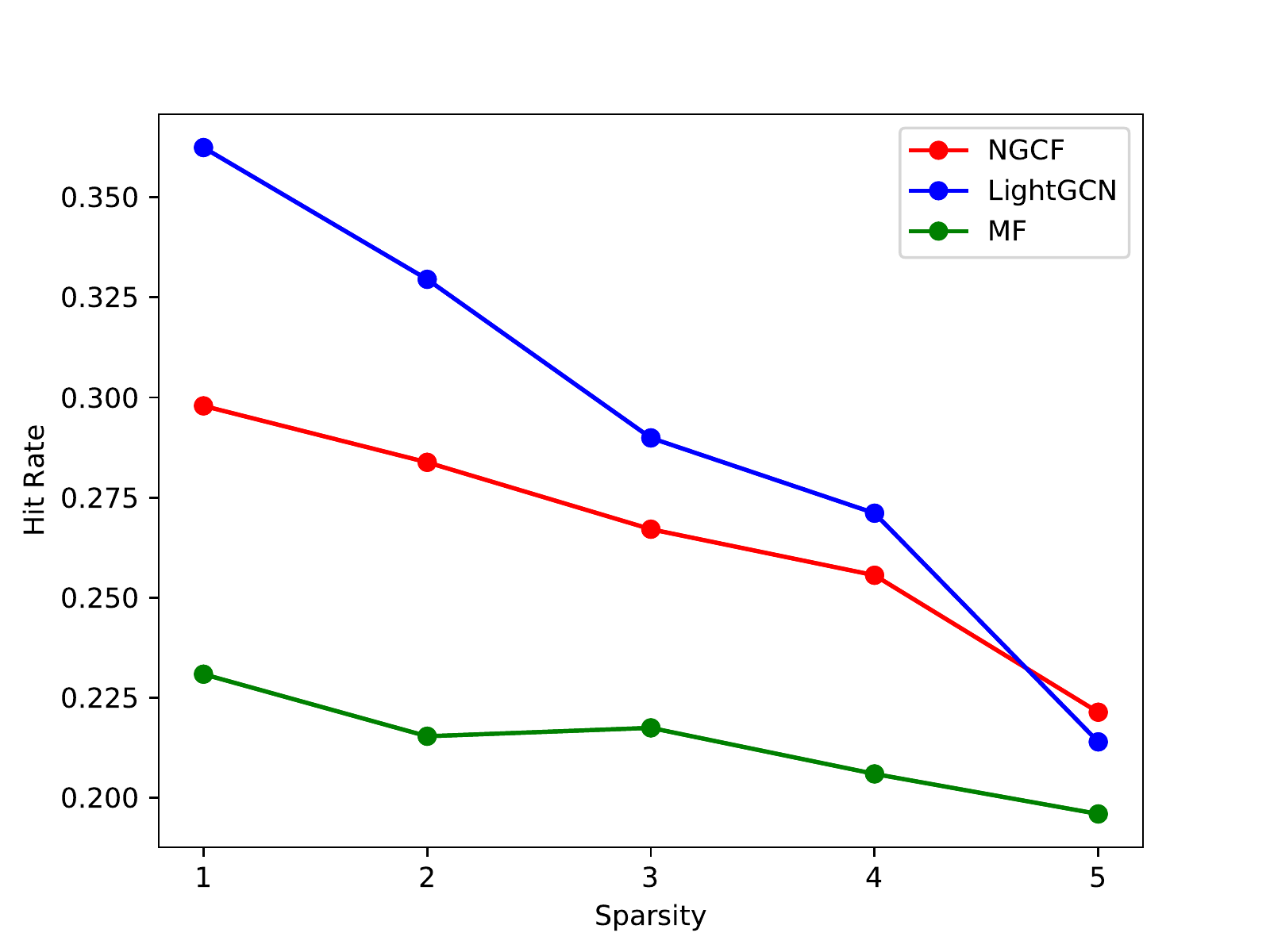} }}
    \subfloat[Tianchi-4527720\label{fig:intro-45}]{{\includegraphics[width=0.5\linewidth]{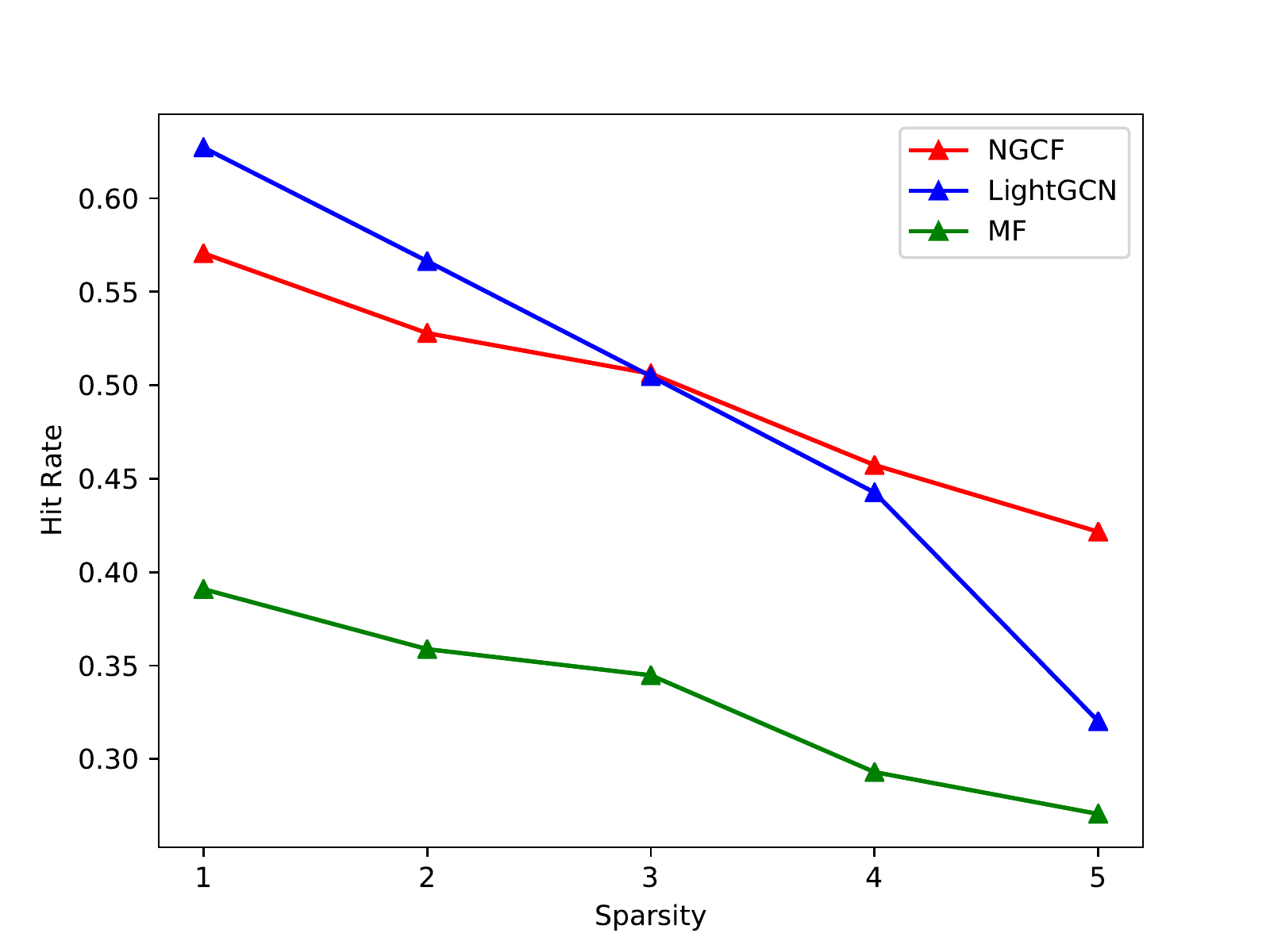} }}
	
\vspace{-0.1in}
\caption{The Performance of MF, NGCF and LightGCN vs. the Graph Sparsity on Two Tianchi datasets.}\label{fig:intro}
\vspace{-0.3in}
\end{figure}

The key advantage of GNN-based CF methods is to learn both the embeddings and a GNN model by explicitly capturing the structural information in the user-item interaction graph. However, we empirically found that the improvement may no longer exist when the bipartite graph is very sparse.  In Figure~\ref{fig:intro}, we illustrate how the sparsity of user-item bipartite graph affects the performance of representative CF models including MF, NGCF and LightGCN. Specifically, we adjust the sparsity of the user-item graphs from two different commodity categories in the Tianchi dataset~\cite{tianchi} by randomly removing edges from the original graph while not disrupting the graph connectivity. We use $x\in\{1,2,3,4,5\}$ (i.e., the x-axis) to denote the sparsity of a bipartite graph where the larger $x$ is, the sparser the graph is. As shown in Figure~\ref{fig:intro}, the performance of all the methods drops when the graph sparsity increases, in which GNN-based methods (NGCF and LightGCN) yield more dramatic performance reduction compared with MF. The inferior performance of GNN-based methods under the sparse setting could be attributed to the difficulty to learn high-quality user/item embeddings from limited user-item interactions.
However, the user-item interaction graphs are often sparse in the real-world recommendation scenarios~\cite{adomavicius2005toward}. Thus, in this paper, we study how to alleviate the challenge of sparsity while enjoying the merits of high-order topological connectivity. Note that in this work, we focus on addressing the recommendation tasks solely based on the historical interactions following the previous works~\cite{wang2019neural,he2020lightgcn}. Our motivation lies in learning CF signals from local structures of the user-item interaction graph, and the proposed model is called Localized Graph Collaborative Filtering (LGCF). 
Particularly, given a user and an item, LGCF makes predictions about their interaction based on their local structural context in the bipartite graph, rather than the user and item embeddings. Further comparisons and discussions between the traditional GNN-based CF methods and the proposed LGCF can be found in~\textbf{Appendix A}.

To achieve the goal of LGCF, we face two main challenges: (1) how to construct the localized graph given two target nodes, and (2) how to capture the important CF information from the localized graph. To solve these two challenges,  LGCF first samples a set of nodes for two target nodes from their neighborhoods, and then generates a localized graph with this node set and their edges. Next, LGCF adopts a GNN model to extract a graph representation for each localized graph. To better capture the graph topological information, LGCF generates a label for each node according to their topological importance in the localized graph, and takes the set of node labels as input attributes. Finally, LGCF makes predictions about the target user and item based on their corresponding localized graph representation. 

The contributions of this paper are summarized as follows: (1) We study a new perspective to build GNN-based CF models for recommendations. Specifically, we aim at learning the recommendation related patterns in the localized graphs induced from the input bipartite graph, instead of learning user and item embeddings; (2) We propose Localized Graph Collaborative Filtering (LGCF) which utilizes a GNN model and minimum distances based node labeling function to generate graph representation for each user-item pair. In this way, we can encode the localized graph topological information and high order connectivity into the graph representation simultaneously; (3) We validate the effectiveness of LGCF on numerous real-world datasets. Especially, LGCF can achieve significantly better performance than representative GNN-based CF methods when user-item bipartite graphs are sparse. 
\begin{figure*}[t]
    \centering
    \includegraphics[width=0.8\linewidth]{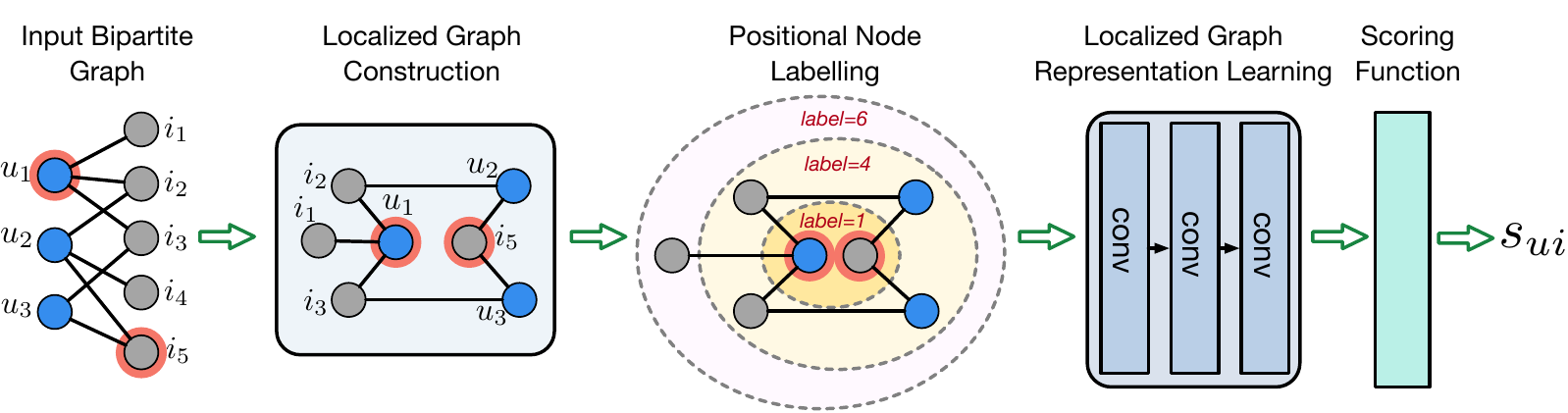}
    \vskip -1em
    \caption{An Overview of the Proposed Framework LGCF.}
    \label{fig:frame}
    \vskip -1em
\end{figure*}

\section{Problem Statement.}\label{sec:problemstatement}
A CF-based recommendation dataset can be formulated as a bipartite graph $G=({\bf U},{\bf I},{\bf E})$, in which ${\bf U} = \{u_1,u_2,...,u_n\}$ denotes a set of $n$ users, ${\bf I} = \{i_1,i_2,...,i_m\}$ is the set of $m$ items, and $\bf E=\{e_0,e_1,...,e_l\}$ (e.g., $e_j=(u_{e_j},i_{e_j})$) indicates the edge set describing the historical interactions between users and items. Following previous works~\cite{wang2019neural,he2020lightgcn}, we assume that users and items have no content information. 
Given a candidate pair $(u_j, i_{k})$ consisting of a target user $u_{j}$ and a potential item $i_{k}$, we need to calculate a preference score $s_{j,k}$ to indicate how likely this potential item should be recommended to the target user.  
To solve this problem, we aim at learning a desirable score mapping function $f$, which generates a preference score $s_{jk}=f(u_j,i_k | \textbf{E})$ to make recommendations based on the historical interactions $\textbf{E}$.

\section{The Proposed Framework.}\label{sec:model}

In this section, we first give an overall description of the proposed framework, then detail its key modules, next introduce the optimization objective and finally discuss integration strategies with existing GNN-based CF methods. Note that compared to existing GNN-based recommendation methods, the mode size of LGCF is much smaller and is independent on the number of users and items. More details about the mode size analysis can be found in~\textbf{Appendix B}.

The goal of LGCF is to learn CF-related knowledge for each user-item pair from its local structures in an interaction bipartite graph. To achieve this goal, it provides solutions to tackle two aforementioned obstacles -- a local structure extraction to construct the localized graph for the user-item pair and a powerful model to capture the CF-related patterns from the localized graph.
An overview of LGCF is illustrated in Figure~\ref{fig:frame}. Specifically, given the input bipartite graph, LGCF first constructs the localized graphs centered with the target user and the target item. To preserve edges with rich CF information into the localized graph, we develop a localized graph construction method to efficiently generate the localized graphs. After that, LGCF labels the nodes within the localized graphs according to the topological positions relative to the target nodes. Then a GNN-based graph representation learning module will embed the high-order connectivity along with the node positional annotations simultaneously.
Through this process, LGCF eliminates the traditional embedding-based strategy and effectively captures the critical CF-related patterns. Finally, the scoring function module calculates the preference score based on the learned graph representation. To sum up, there are three key modules in this framework: (1) \textbf{the localized graph construction} module that extracts a localized graph covering most edges related to a given user-item pair; (2)  \textbf{the localized graph representation learning module} that learns recommendation related representations from the localized graph ; and (3) \textbf{ the scoring function module }that computes a preference score based on the learned graph representation. Next we will detail each module.

\subsection{Localized Graph Construction.}
\begin{figure*}[t]
    \centering
    \includegraphics[width=0.7\linewidth]{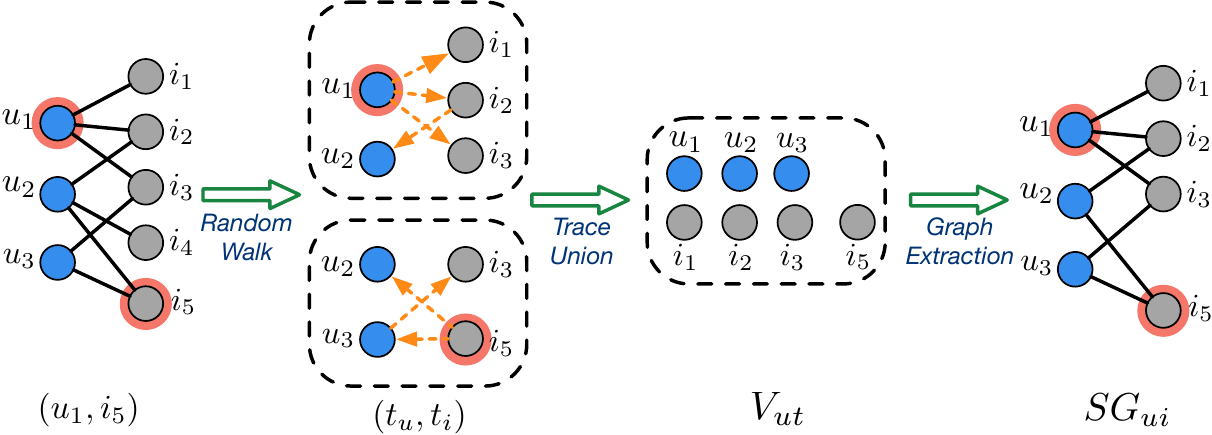}
    \vskip -1em
    \caption{An Illustration of Localized Graph Extraction.}
    \label{fig:frameLSE}
    \vskip -1em
\end{figure*}

The localized graph construction module is designed to extract the localized graph covering the most important edges (i.e., collaborative filtering information) for a given user-item pair. We need to extract a localized graph for each user-item pair in the training process and the inference process, thus we aim to include the most representative edges for each pair with the consideration of scalability. To achieve these goals,  we propose a localized graph construction as shown in Figure~\ref{fig:frameLSE}. This framework can be divided into three steps: (1) \textbf{step 1 -random walk}: we first take advantage of random walk with restart (RWR)~\cite{tong2006fast} method to sample neighboring nodes for the user node and the item node. By using RWR, we can sample neighboring nodes which are different hops away from the center node. As a consequence, we can get a representative localized graph which captures abundant context information. Meanwhile, RWR is a sampling method in essence, thus it can naturally improve the scalability of the proposed LGCF.
Specifically, given a user-item bipartite graph $G$ and a user-item pair $(u,i)$, we perform a random walk on $G$ from the user node $u$ and the item node $i$, respectively. Starting from the original node, each walk travels to one of the connected nodes iteratively with a uniform probability. In addition, there is a pre-defined probability for each walk to return to the starting node at each step. In this way, more neighboring nodes can be included in the walk. Via RWR, we can get two traces $t_u$ and $t_i$ for the user node $u$ and the item node $i$, respectively. Each trace then can be denoted as a node subset, i.e., $V_u$ and $V_i$. (2) \textbf{step 2 - trace union}: Next, we merge $V_u$ and $V_i$ into a union node subset $V_{ui}$. (3) \textbf{step 3 - graph extraction}: With this union node subset $V_{ui}$, we can extract a localized graph from the original user-item graph. Specifically, all the nodes in $V_{ui}$ and all the edges among these nodes are extracted from the original graph to build a localized graph, which is denoted as $SG_{ui}$. To conclude, $SG_{ui}$ consists of neighboring nodes of different hops away from $u$ and $i$ and thus preserves important collaborative filtering information for the given user-item pair $(u,i)$.

\subsection{Localized Graph Representation Learning.}
This module aims at learning an overall representation for each user-item pair based on its extracted localized graph. We expect to encode the CF information and also the high-order connectivity related to a user-item pair into this representation. Therefore, we propose to use a graph neural network model since it can naturally capture the structural information and explicitly encode the high-order connectivity into the representation. Also, to further encode the topological information, we adopt a node labeling method to generate a positional label for each node based on its distance towards the user-item pair~\cite{zhang2018link} where the node labels are considered as the input graph node attributes. Next, we will introduce the node labeling method and the adopted GNN model.

\subsubsection{The Node Labeling Method.}
Following the common setting from the previous works~\cite{wang2019neural,he2020lightgcn}, we do not assume that node attributes are available for LGCF. Instead, we generate a label for each node to indicate its unique role in the localized graph. Specifically, there are three expectations for the generated label. First, it is important to distinguish the target user and item from other nodes through the generated node labels, as the model should be aware of the target user-item pair.  Second, for contextual nodes excluding the target user and item, they play different roles in the recommendation prediction considering their different relative position towards the target user and item. Thus, we need to distinguish these nodes via the generated node labels based on their positions; Finally, for the consistency of node labels, the closer the node is to both the target user and target item, the similar its label should be with that of the target ones. Therefore, we propose to use a node labeling method based on its distance to the user-item pair via Double-Radius Node Labeling (DRNL)~\cite{zhang2018link} to generate a label for each node, and take the generated node labels as the input node attribute for the GNN model.
The key motivations of DRNL are to distinguish the target user node and the item node from other nodes while preserving their relatively positional relations. Specifically, we first assign label $1$ to the target user node and the target item node to distinguish them from other nodes. Next, we assign labels to other nodes based on their minimum distances towards two target nodes on the extracted localized graph. Intuitively the closer the user or item node is to the target nodes, the more similar its label should be with that of the target nodes. For a specific node $x$ on the graph, we evaluate its distance towards the target user and the target item by summing up its minimum distances to these two nodes. Since we set the label of the target user and the target item as $1$, we will also assign these nearby nodes labels with small values. In particular, given nodes $x$ and $y$, if the distance between $x$ and the target nodes is smaller than that of $y$, the label value of $x$ is supposed to be smaller than that of $y$. If the distances are the same, the node with a smaller minimum distance to the target user or the target item should have a label with a smaller value. DRNL adopts a hashing function $f_l()$ satisfying the criteria described above to compute node labels. The node labeling function $f_l(x)$ for each node $x$ is summarized as follows:
\begin{align}
f_l(x)=
\begin{cases}
1,& \text{x=u or x=i,}\\
1+\min(d_u,d_i)+(d/2)^2,& \text{otherwise,}
\end{cases}
\end{align}
where $u$ and $i$ denote the target user node and item node, respectively. $d_u$ ($d_i$) represents the minimum distance between nodes $x$ and $u$ ($i$). $d=(d_u+d_i)$ is the sum of minimum distances which should be an odd due to the nature of the bipartite graphs. 

\subsubsection{The GNN Model.}
For each user-item pair$(u,i)$, we have extracted its localized graph $SG_{ui}$, and it can be denoted as $SG_{ui}=\{{\bf A}_{ui},{\bf X}_{ui}\}$, where ${\bf A}_{ui}$ is the adjacency matrix of $SG_{ui}$, and ${\bf X}_{ui}$ is the input node attributes generated by above node labeling method. It is natural to take advantage of a GNN model to learn an overall representation for each pair based on its localized graph $SG_{ui}$, since GNNs have shown great potential in capturing complex patterns on graphs~\cite{xu2018powerful}. There are typically two key operations in a GNN model -- the graph filtering operation to refine node representations and the graph pooling operation to abstract the graph-level representation from the node representations. In this work, we adopt graph convolutional filtering~\cite{kipf2016semi} as the graph filtering operation. Its process can be described as follows:

\begin{equation}
{\bf X}_{l+1} = \sigma(\tilde{\bf D}^{-\frac{1}{2}}\tilde{\bf A}\tilde{\bf D}^{-\frac{1}{2}}{\bf X}_l{\bf W}), 
\label{eq: filter_gcn}
\end{equation}
where ${\bf X}_l$ denotes the node representations in the $l$-th GNN layer, $\bf W$ is the transformation matrix to be learned, $\tilde{\bf A} = {\bf A}_{ui} + {\bf I}$ represents the adjacency matrix with self-loops,  $\tilde{\bf D} = \sum_{j}\tilde{\bf A}_{jj}$ is the diagonal degree matrix of $\tilde{\bf A}$ and $\sigma()$ is the activation function. The node labels are set to be the input node representations ${\bf X}_0$. The graph pooling operation adopted by LGCF is sum pooling, which sums up representations all over the nodes for the graph-level representation. Its process can be denoted as follows:

\begin{align}
    {\bf x}_{ui}  = pool({\bf X}_{L}) = sum({\bf X}_{L}),
\label{eq: pool_gcn}
\end{align}
where ${\bf x}_{ui}$ denotes the graph representation for the user-item pair $(u,i)$ and ${\bf X}_{L}$ is the node representations outputted by the last layer. The overall process of the GNN model can be summarized as:

\begin{align}
    {\bf x}_{ui} = GNN({\bf A}_{ui}, {\bf X}_0 ; {\bm \theta}_{GNN}),
\end{align}

where ${\bm \theta}_{GNN}$ denotes the parameters to be learned in the GNN model. It is straightforward to extend LGCF with other graph filtering operations~\cite{xu2018powerful,velivckovic2017graph} and graph pooling operations~\cite{ma2019graph,gao2019graph}, and we will leave such investigation as one future work.

\subsection{The Scoring Function.}

The scoring function aims at giving a score for a user-item pair $(u,i)$ based on its corresponding localized graph representation ${\bf x}_{ui}$. The higher the score is, the more likely the item should be recommended to the user. We use a single-layer linear neural network to model the scoring function in LGCF, which can be described as follows:

\begin{align}
     s_{ui} = score({\bf x}_{ui}) = \sigma({\bf x}_{ui}^T*{\bf w}),
\label{eq:score}
\end{align}
where $\bf w$ is the transformation vector to be learned and $\sigma$ is the sigmoid function. 

\subsection{Model Optimization.}
In this work, we adopt the pairwise Bayesian Personalized Ranking (BPR) loss to train the framework. BPR loss is one of the most popular objective functions in recommendation tasks. It takes the relative order of the positive node pairs and negative node pairs where the positive node pairs are existent user-item node pairs observed in graphs, while the negative node pairs are non-existent user-item node pairs generated by negative sampling. Specifically, BPR assumes that the prediction scores of the positive node pairs should be larger than these of the corresponding negative ones. The objective function of our model is as follows:
\begin{align}
    \min_{{\bm \Theta}_{GNN},{\bm w}}\sum_{(u,i,i^{'})\in \mathcal{O}} -\ln\sigma(s_{ui}-s_{ui^{'}}),\label{eq:BPR-loss}
\end{align}
where $\mathcal{O}= \{(u,i,i^{'})|(u,i)\in E, (u,i^{'})\in E^-\}$ denotes the training data for the graph $G$, $E$ is the set of the existent interactions between users and items in $G$ and $E^-$ indicates the set of  the non-existent interactions. In this work, we adopt ADAM~\cite{kingma2014adam} to optimize the objective.

\subsection{Integration with Embedding-based CF Methods.}

LGCF aims at capturing the collaborative filtering information from a localized perspective by encoding the localized CF information into a subgraph level representation, while embedding-based GNN methods such as NGCF and LightGCN are to learn a set of user and item embeddings in the latent space based on all the user-item interactions. Therefore, LGCF provides a new perspective to build GNN based recommendations. As a consequence, LGCF is likely to provide complementary information to existing GNN based methods and integrating it with existing methods has the potential to boost the recommendation performance. In this subsection, we discuss strategies to combine LGCF with existing GNN based methods. In particular, we propose two strategies, i.e., \textit{LGCF-emb} and \textit{LGCF-ens}. Before we detail these strategies, we first denote the refined user $u$ and item $i$ embedding generated by embedding-based GNN methods as ${\bf h}_u$ and ${\bf h}_i$, and the representation outputted by LGCG for a user-item pair $(u,i)$ as ${\bf x}_{ui}$. In this work, we focus on  how to integrate LGCF with LightGCN given the competitive performance of LightGCN.

\subsubsection{LGCF-emb.}
In LGCF-emb, we propose to directly concatenate the embeddings generated by LGCF and LightGCN, and then the scoring function takes the concatenated embedding as input to generate a score. Finally, the LGCF model and the LightGCN model are jointly trained by optimizing the objective function in Eq.~\ref{eq:BPR-loss}. The concatenation process is summarized as follows:
\begin{align}
   {\bf h}_{ui}=({\bf h}_u*{\bf h}_i)||{\bf x}_{ui},\label{eq:LGCF-emb}
\end{align}
where $*$ denotes element-wise multiplication and $||$ represents the concatenation operation. 

\subsubsection{LGCF-ens.}
In LGCF-ens, we propose to combine LGCF and LightGCN in an ensemble way. Specifically, we conduct a weighted combination of the scores generated by LGCF and LightGCN. In addition, instead of joint training, we train LGCF and LightGCN, separately. The final score for a user-item pair $(u,i)$ can be computed as follows:
\begin{align}
    s_{ui}=score({\bf x}_{ui}) + \lambda \cdot ({\bf h}_u^\mathrm{T}{\bf h}_i),\label{eq:LGCF-ens}
\end{align}
where $score()$ denotes the scoring function in Eq~\ref{eq:score} and $\lambda$ can be either a predefined hyper-parameter or learnable parameter.

\section{Experiment.}\label{sec:experiments}

In this section, we conduct extensive experiments on various real-world datasets to validate the effectiveness of the proposed LGCF. Via experiments, we aim to answer the following questions: \textbf{RQ1}: How does LGCF perform compared with the state-of-the-art CF models on the sparse setting? \textbf{RQ2}: How does LGCF perform compared with the state-of-the-art CF models on the normal setting? \textbf{RQ3}: Can LGCF be complementary to the embedding-based GNN CF models? \textbf{RQ4}: How does the sparsity of data affect the performance of LGCF and other CF models?

In the following subsections, we will first introduce experimental settings, next design experiments to answer the above questions and we further probe to understand the working of LGCF. 

\subsection{Experimental Settings.}

\subsubsection{Datasets.}
In our experiments, we have tested the proposed framework on twelve datasets, which are from three different real-world recommendation scenarios: Tianchi\cite{tianchi}, Amazon\cite{amazon} and MovieLens\cite{harper2015movielens}. Specifically, we construct seven datasets corresponding to seven item categories from Tianchi and denote each of them as Tianchi-$ID$, where $ID$ indicates the item category, two datasets from two item categories from Amazon and two datasets from two item categories from MovieLens. 
More details about these datasets can be found in the {\bf Appendix C}. Note that it is common to extract a user-item interaction dataset based on the item category from a large recommendation dataset for evaluations that has been widely adopted by existing works~\cite{amazon,wang2019neural,he2020lightgcn}.




\subsubsection{Baselines and Evaluation Metrics.} We mainly compare the proposed method with LightGCN~\cite{he2020lightgcn} and NGCF~\cite{wang2019neural}, which have empirically outperformed most recent CF-based models including NeuMF~\cite{he2017neural}, PinSage~\cite{ying2018graph} and HOP-Rec~\cite{yang2018hop}. In addition, we also compare LGCF with one of the most classic factorization-based recommendation models -- MF~\cite{koren2009matrix}. We use HR and NDCG as evaluation metrics in our experiments. More details about the baselines and evaluation details can be found in  {\bf Appendix D}.

\begin{table*}

 \caption{Densities of the Sparse User-item Bipartite Graphs. Note that ``Density" is computed specifically for the interaction graph as $Density=\#Edges\div(\#Users\times\#Items)$.}
 \vspace{-0.15in}
  \centering
  \scalebox{0.85}{
  \begin{tabular}{c|ccccccc|cc|cc}
  \hline
  \multirow{2}{*}{}&\multicolumn{7}{c}{\textbf{Tianchi}}&\multicolumn{2}{c}{\textbf{Amazon}} &\multicolumn{2}{c}{\textbf{MovieLens}}\\ \cline{2-12}
  
                 &\textbf{685988}  &\textbf{2798696} &\textbf{4527720} &\textbf{174490}   &\textbf{61626} & \textbf{810632} & \textbf{3937919} &\textbf{Beauty} &\textbf{Gift} &\textbf{War} &\textbf{Romance}\\ \hline
 density & 0.0019   & 0.0106  &0.0074 & 0.0036   & 0.0014 & 0.0019 & 0.0064 &0.0121 &0.0071&0.0113&0.0027               \\

\hline
  \end{tabular}}
  
\vspace{-0.1in}
 \label{table:sparsity-graph-den}
\end{table*}


\subsection{Performance in Sparse Scenarios.}
The major motivation of LGCF is to improve the recommendation performance when there are a few user-item interactions or user-item interactions are sparse. Therefore, we start the evaluation by comparing the proposed method with baselines under the sparse setting, which correspondingly answers \textbf{RQ1}. Specifically, to obtain a sparse training user-item interaction bipartite graph, we randomly select as many interactions as possible as the validation set or the test set while ensuring no isolated nodes in the training set and no cold-start nodes in the validation set and the test set. After this process, the densities of the resulting sparse graphs are summarized in Table~\ref{table:sparsity-graph-den}.

\begin{table*}
 \caption{Performance Comparison under the Sparse Setting.}
 \vspace{-0.15in}
  \centering
  \scalebox{0.8}{
  \begin{tabular}{c|ccccccc|cc|cc}
  \hline
  \multirow{2}{*}{HR($\%$)}&\multicolumn{7}{c}{\textbf{Tianchi}}&\multicolumn{2}{c}{\textbf{Amazon}} &\multicolumn{2}{c}{\textbf{MovieLens}}\\ \cline{2-12}
  
                 &\textbf{685988}  &\textbf{2798696} &\textbf{4527720} &\textbf{174490}   &\textbf{61626} & \textbf{810632} & \textbf{3937919} &\textbf{Beauty} &\textbf{Gift} &\textbf{War} &\textbf{Romance}\\ \hline
MF & 28.4$\pm$2.3 & 19.8$\pm$4.9 &25.4$\pm$3.1 &27.0$\pm$1.7  & 15.6$\pm$1.2 &25.2$\pm$1.9  & 38.7$\pm$5.2 & 64.3$\pm$7.4 & 12.7$\pm$0.3 & 55.8$\pm$3.0 & 57.7$\pm$1.0              \\
NGCF & 36.7$\pm$3.0 & 25.6$\pm$4.2 & 32.7$\pm$3.3 &37.2$\pm$4.2  & 17.8$\pm$2.2  & 36.0 $\pm$2.3  &47.5$\pm$3.0 &83.1$\pm$0.7 &14.4$\pm$0.3 &63.5$\pm$1.1 &  71.0$\pm$0.2             \\            
LightGCN  & 29.4$\pm$2.1  & 22.1$\pm$3.9 & 25.9$\pm$ 3.6 & 25.9$\pm$3.6  & 18.8$\pm$1.8 & 28.2$\pm$1.4 &24.4$\pm$1.7 & 39.1$\pm$6.6&14.5$\pm$0.3& 32.4$\pm$2.0&36.0$\pm$1.0               \\            
\hline
LGCF  & 65.3$\pm$0.4  & 39.7$\pm$0.6 &  60.2$\pm$1.0 & 66.2$\pm$0.5 & 41.6$\pm$3.0  & 60.7$\pm$0.2  &61.6$\pm$0.2 &88.0$\pm$0.4 &32.4$\pm$2.0 &81.1$\pm$0.5 & 82.7$\pm$ 1.2             \\        
\hline
  \end{tabular}}
 \label{table:sparsity-graph-per}
\end{table*}

The performance comparison on these sparse datasets is demonstrated in Table~\ref{table:sparsity-graph-per}. We make the following observations: 1) The performance of LightGCN is worse than NGCF on most datasets. These results suggest that only neighborhood aggregation may not help embedding refinement in sparse scenarios; 2) LGCF outperforms other baselines on the sparse datasets significantly, which supports that capturing local structures is more effective than learning user and item embeddings in sparse scenarios. Meanwhile, this validates the effectiveness of LGCF in capturing CF information from local structures on sparse user-item bipartite graphs. More details about how the sparsity impacts the performance and why LGCF can improve the performance will be provided in the following subsections. 

\subsection{Performance in Normal Scenarios.}
In the last subsection, we empirically demonstrated the superior of LGCF over the baseline methods in sparse scenarios. In this section, we further study how LGCF performs in normal settings where we ensure we have sufficient training data. To achieve the goal, we randomly select 90\% of user-item interactions as the training set, and then divide the remaining interactions into the validation set and the test set equally. Since LGCF makes recommendations from a different perspective from LGCF and LightGCN, we also investigate if LGCF is complementary to them. These experiments aim to answer \textbf{RQ2} and \textbf{RQ3}. The results are summarized in Table~\ref{table:sparsity-graph-per}. It can be observed: 1) LGCF can often perform reasonably that achieves comparable performance with the baseline methods on most datasets; 2) LGCF-ens has shown significant performance improvement over both LGCF and LightGCN on most datasets. This demonstrates that LGCF indeed provides complementary information to embedding-based CF methods. Thus, there is a great potential to integrate these methods for the recommendation task; 3) LGCF-emb does not perform well on some datasets. It directly concatenates embeddings from LGCF and LightGCN. However, the embeddings from LGCF and LightGCN could be not aligned. Among the integration methods, LGCF-ens achieves consistent and better performance compared to LGCF-emb.

\begin{table*}
 \caption{Performance Comparison under the Normal Settings.}
 \vspace{-0.15in}
  \centering
  \scalebox{0.8}{
  \begin{tabular}{c|ccccccc|cc|cc}
  \hline
  \multirow{2}{*}{HR($\%$)}&\multicolumn{7}{c}{\textbf{Tianchi}}&\multicolumn{2}{c}{\textbf{Amazon}} &\multicolumn{2}{c}{\textbf{MovieLens}}\\ \cline{2-12}

                 &\textbf{685988}  &\textbf{2798696}  &\textbf{4527720} &\textbf{174490}   &\textbf{61626} & \textbf{810632} & \textbf{3937919} &\textbf{Beauty} &\textbf{Gift} &\textbf{War} &\textbf{Romance}\\ \hline
MF & 57.3$\pm$3.0 & 22.0$\pm$3.0 & 43.9$\pm$4.1 &41.6$\pm$3.3  & 22.3$\pm$4.6 & 70.7$\pm$ 1.6 & 38.7$\pm$5.3 & 93.7$\pm$1.4 & 44.2$\pm$4.1 & 72.0$\pm$1.7 & 77.7$\pm$0.5            \\
NGCF  & 69.6$\pm$1.8 & 29.7$\pm$3.7 & 56.0$\pm$2.1 & 55.2$\pm$3.0 & 37.1$\pm$3.7 & 53.6$\pm$2.9  & 47.5$\pm$3.0 & 93.5$\pm$1.1 & 46.4$\pm$2.4 & 75.5$\pm$0.5 & 82.6$\pm$0.2            \\            
LightGCN  & 76.2$\pm$1.2 & 32.4$\pm$2.9 &62.7$\pm$1.7 & 60.3$\pm$1.1 & 44.0$\pm$3.5 & 74.1$\pm$0.6  & 55.6$\pm$3.6 & 94.4$\pm$1.1 & 51.4$\pm$0.3& 80.9$\pm$1.0 & 85.3$\pm$0.2             \\            
\hline
LGCF & 76.9$\pm$1.2  & 49.3$\pm$2.5 & 59.3$\pm$1.7 & 63.4$\pm$3.5 & 46.9$\pm$5.6 & 71.2$\pm$2.3  & 55.7$\pm$1.6 & 91.5$\pm$2.5 & 36.9$\pm$6.1 & 65.5$\pm$12.9 & 77.3$\pm$5.6              \\        
\hline
LGCF-emb & 72.1$\pm$7.1& 46.3$\pm$3.1& 54.1$\pm$5.9 & 62.4$\pm$3.7  & 50.0$\pm$3.0 & 71.0$\pm$3.1 &  57.8$\pm$3.0 & 92.6$\pm$2.0 & 36.6$\pm$8.3 & 64.7$\pm$7.6 & 81.5$\pm$2.8              \\       
LGCF-ens & 76.8$\pm$1.1  & 44.8$\pm$4.1 & 62.7$\pm$3.3 &67.3 $\pm$2.1& 54.5$\pm$2.4 & 74.9$\pm$1.0 & 61.7$\pm$2.1 & 94.7$\pm$0.7 &52.8$\pm$1.0 & 87.5$\pm$0.4 & 88.5$\pm$0.4          \\    
\hline
  \end{tabular}}
 \label{table:sparsity-graph-per}
\end{table*}

\begin{figure}
  \centering
  \subfloat[Tianchi-2798696\label{fig:sparsity-27}]{{\includegraphics[width=0.5\linewidth]{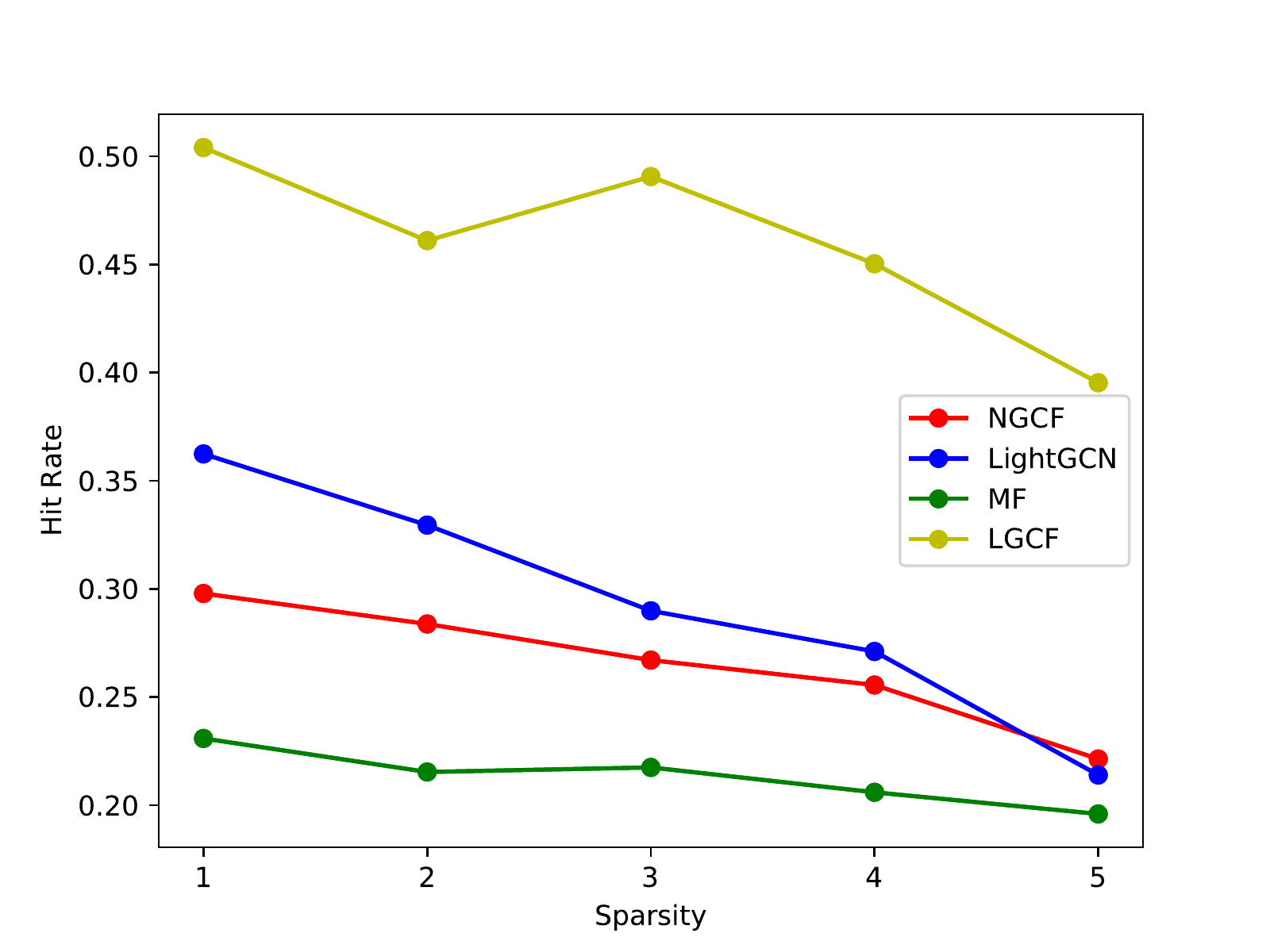} }}
    \subfloat[Tianchi-4527720\label{fig:sparsity-45}]{{\includegraphics[width=0.5\linewidth]{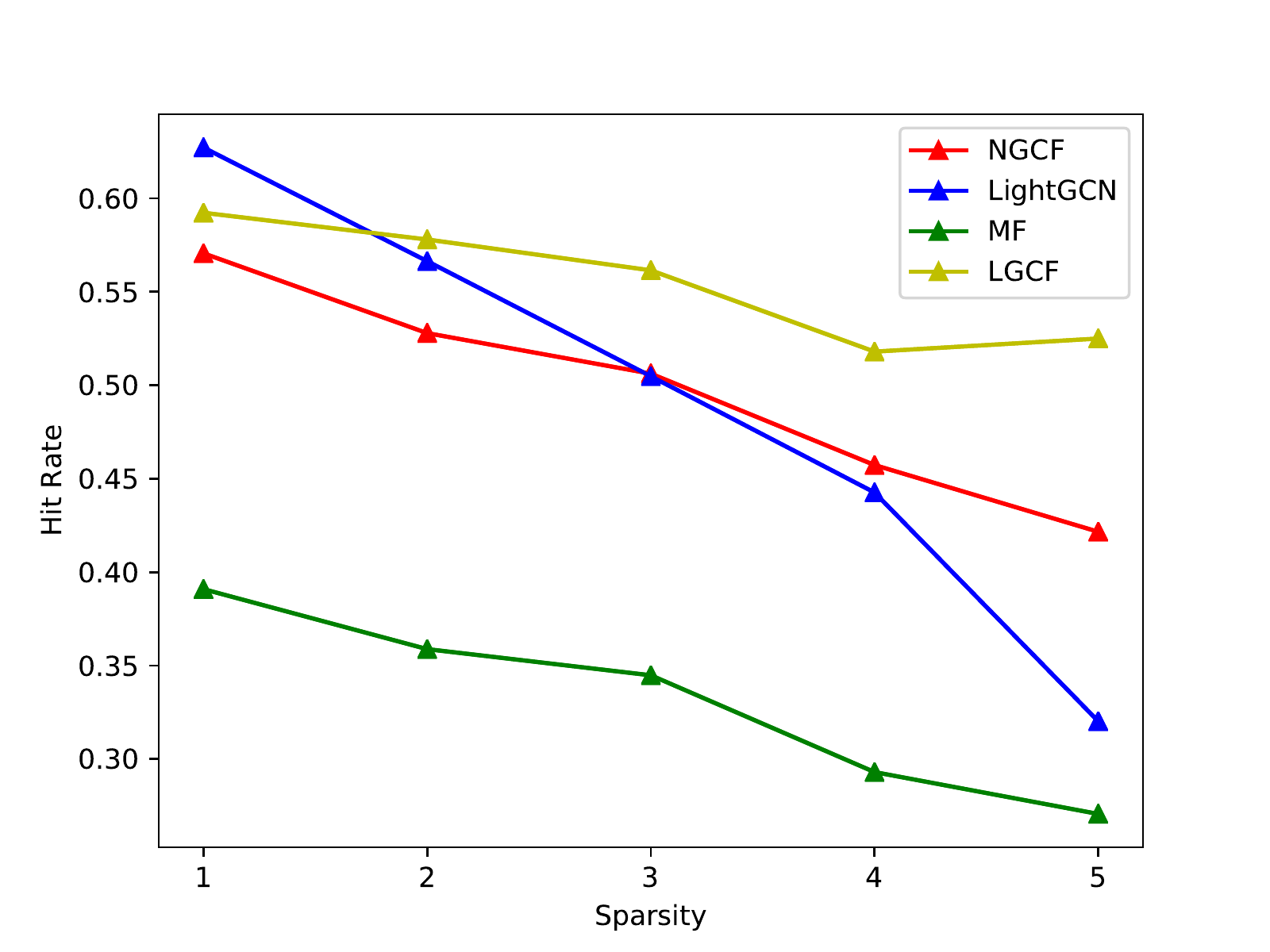} }}
    \quad
    \vspace{-0.1in}
    \subfloat[Tianchi-174490\label{fig:sparsity-17}]{{\includegraphics[width=0.5\linewidth]{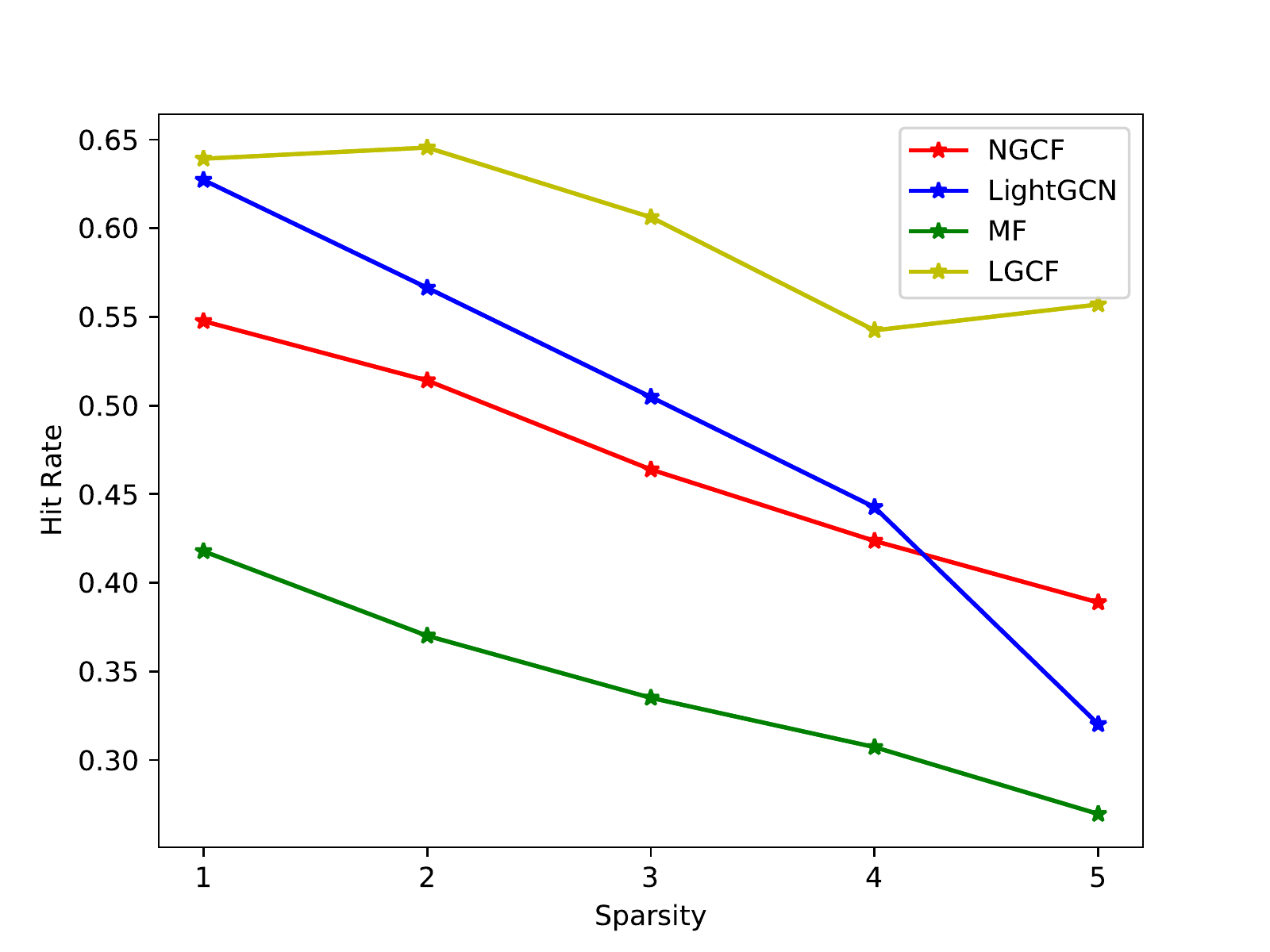} }}
    \subfloat[Tianchi-61626\label{fig:sparsity-61}]{{\includegraphics[width=0.5\linewidth]{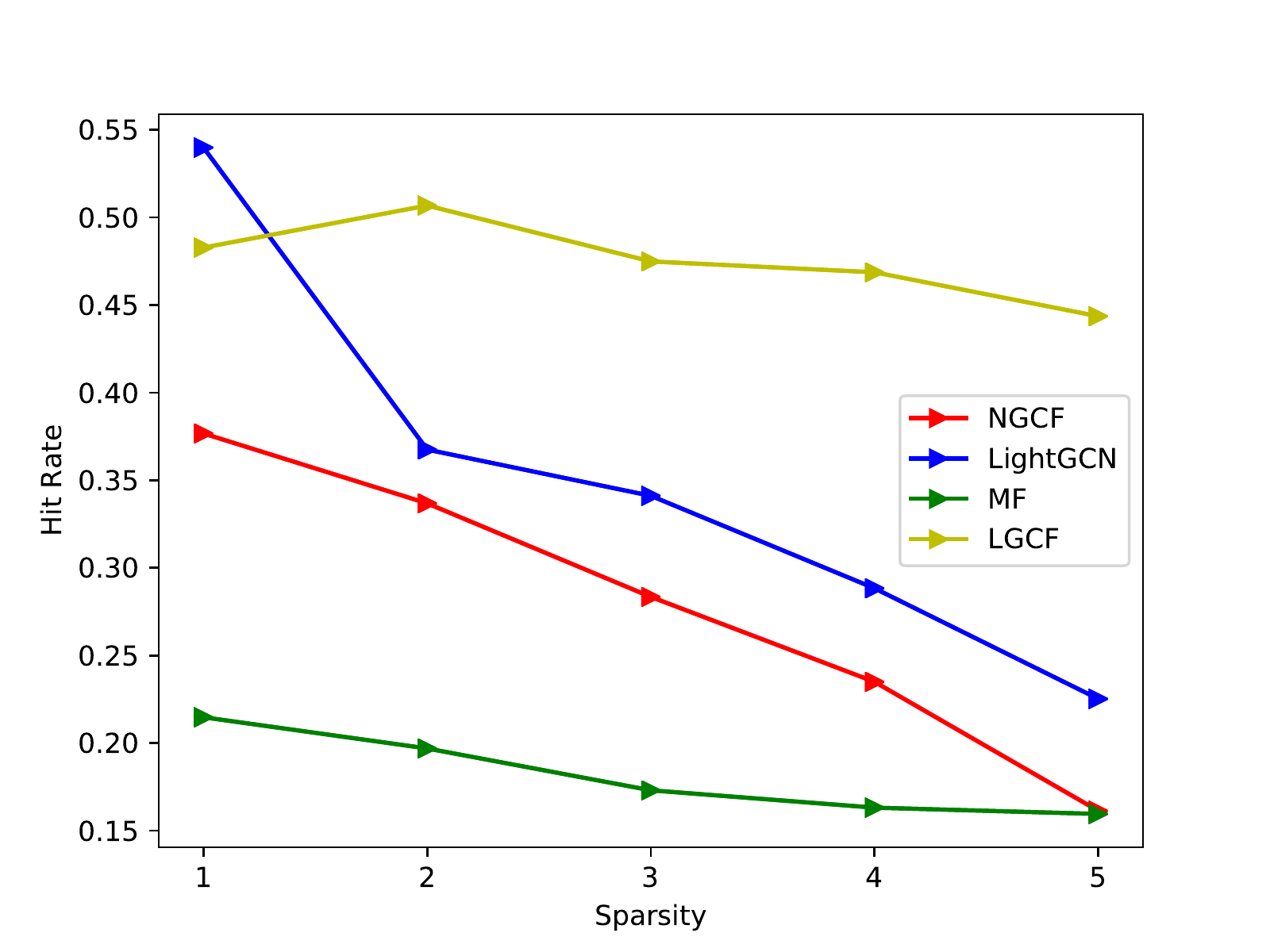} }}

\caption{The Performance of Different Methods vs. the Graph Sparsity on Four Tianchi Datasets }\label{fig:per-vs-sparsity}
\vspace{-0.2in}
\end{figure}

\subsection{The Impact of Sparsity.}
In this subsection, we further explore \textbf{RQ4}: how the sparsity of the training user-item bipartite graph affects the performance of different models. Specifically, we first randomly select 10\%  user-item interactions and split them equally into the validation set and the test set. Note that we constrain there are no cold-start users and items in the validation set and the test set. We take the remaining 90\% interactions as the full training set, and denote its sparsity as $1$. Next, we divide the full training set into two sets: the necessary set and the additional set. The necessary set consists of the minimum interactions to ensure all the users and items connected with at least one neighbor and the additional set includes the remaining interactions. We randomly select 20\%, 40\%, 60\% and 80\% of interactions from the additional set and exclude them from the full training set to increase the sparsity of the training user-item bipartite graphs. Here we denote the sparsity of these training graphs as $2,3,4,5$, respectively. The performance of LGCF, LGCF-enz and the baseline methods are shown in Figure~\ref{fig:per-vs-sparsity}. (We have similar observations on most of datasets, and we select four of them to report due to the space limitation) We can observe that the performance of all the methods tends to decrease with the increase of graph sparsity. However, the performance drop of LightGCN and NGCF is relatively more significant than that of the proposed LGCF. LightGCN and LGCF typically perform comparably on the least sparse cases, and then the performance gap tends to increase with the graph sparsity raises. In addition, LGCF-enz is more robust to the sparsity change compared to LightGCN, which demonstrates that the integration of LGCF can help LightGCN reduce its sensitivity towards graph sparsity.

\subsection{Further Probing.}

In this subsection, we explore to further understand how LGCF works. As mentioned before, LGCF provides a new perspective to build GNN-based CF methods. Experimental results in Table~\ref{table:sparsity-graph-per} have demonstrated that LGCF is complementary to LightGCN. Thus, we first examine how LGCF is complementary to LightGCN for different types of user-item pairs. Moreover, localized graphs play a crucial role in LGCF. Therefore, we investigate if localized graphs for positive and negative pairs are distinguishable via case studies.  

\subsubsection{How is LGCF complementary to LightGCN.} \label{sec:complementary}
\begin{figure}
  \centering
  \subfloat[Tianchi-61626\label{fig:degree-61}]{{\includegraphics[width=0.52\linewidth]{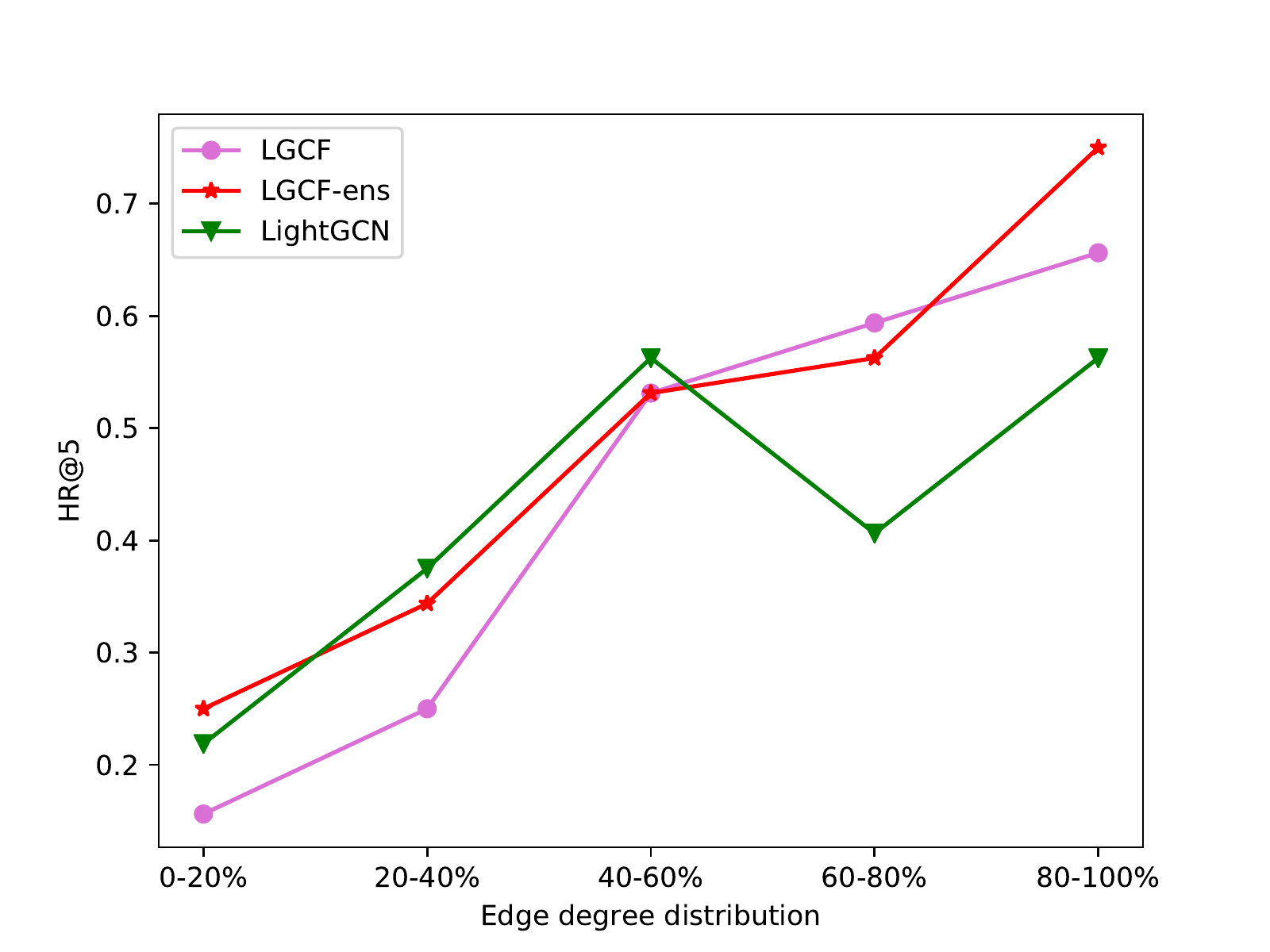} }}
    \subfloat[MovieLens-War\label{fig:degree-war}]{{\includegraphics[width=0.52\linewidth]{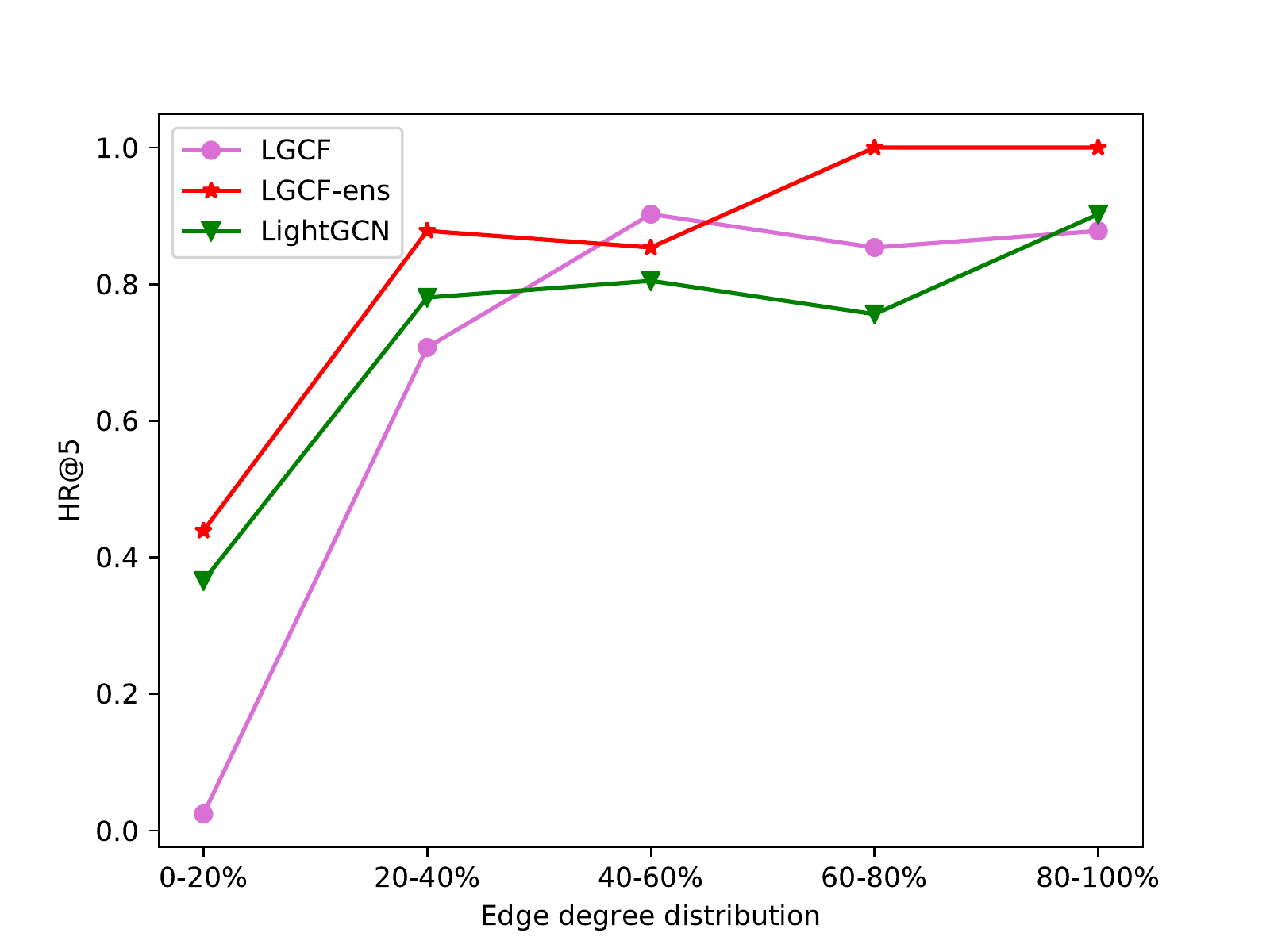} }}

\caption{Performance Analysis with Degree.}\label{fig:per-vs-degree}
\vspace{-0.3in}
\end{figure}

To explore what kind of user-item pairs can benefit by integrating LGCF and LightGCN, we divide the test set in normal scenarios into different groups based on the average degree of the user and the item for a user-item pair. Specifically, we first sort all the user-item pairs in the test set in ascending order according to their average degree, and divide the sorted pairs into $5$ groups. We then train models on the same training set and test them on these groups. The performance of LightGCN, LGCF and LGCF-ens is shown in Figure~\ref{fig:per-vs-degree}. More results on different datasets can be found in {\bf Appendix E}. We first note that the performance of all methods tends to increase when degree increases. In most cases, for all degree groups, LGCF-ens (or integrating LGCF and LightGCN) can boost the recommendation performance. This improvement is relatively more significant for groups with small degree. These results suggest that LGCF provides complementary information to LightGCN for all degree groups especially for these groups with small degree. This property has its significance in practice since how to make recommendations for user-item pairs with limited interactions is still a challenging problem in real-world recommender systems. In addition, this property makes the proposed method potentially computation-efficient for large-scale datasets, since we can first filter items that are less likely to be preferred by a given user based on embeddings and then leverage LGCF to rank the remaining items.

\subsubsection{Localized Graphs.}

\begin{figure}
  \centering
  \subfloat[positive pair$(249,511)$\label{fig:pos-122}]{{\includegraphics[width=0.45\linewidth]{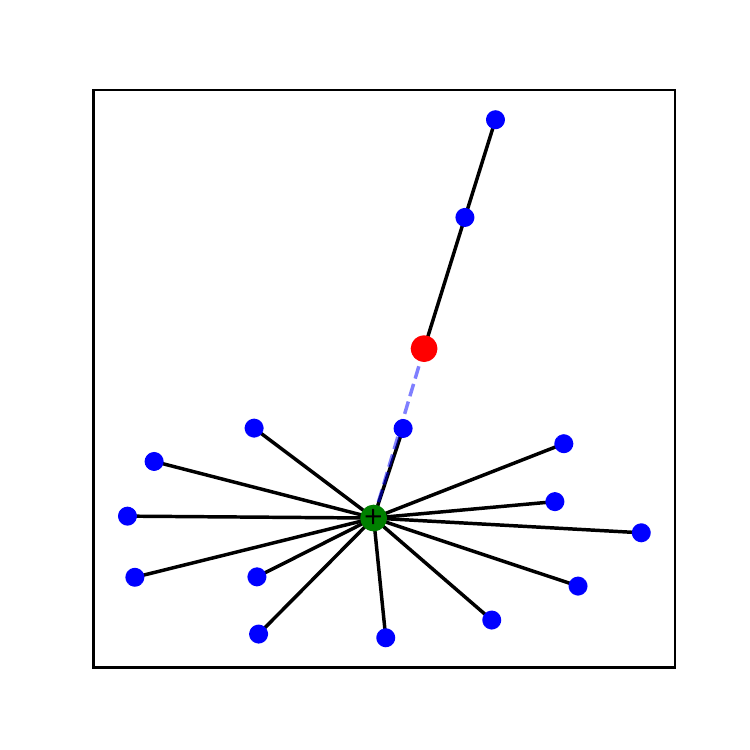} }}
  \subfloat[negative pair$(249,508)$\label{fig:pos-249}]{{\includegraphics[width=0.45\linewidth]{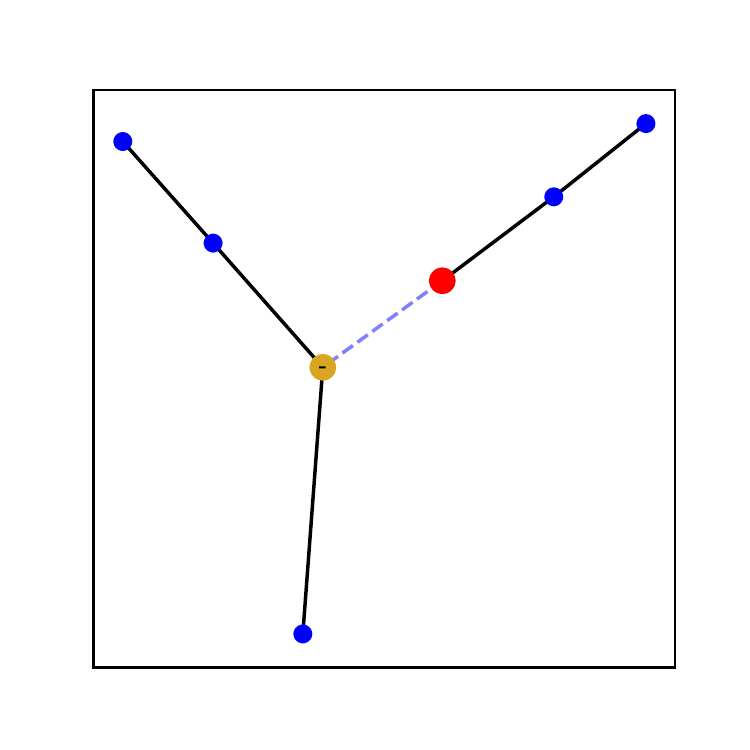} }}

  \quad
  \subfloat[positive pair$(87,459)$\label{fig:neg-122}]{{\includegraphics[width=0.45\linewidth]{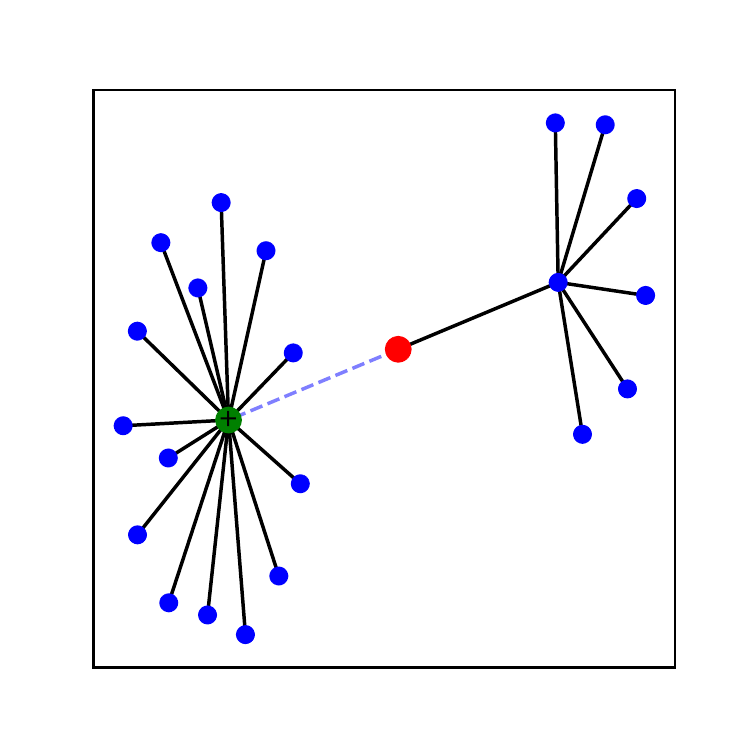} }}
  \subfloat[negative pair$(87,592)$\label{fig:neg-249}]{{\includegraphics[width=0.45\linewidth]{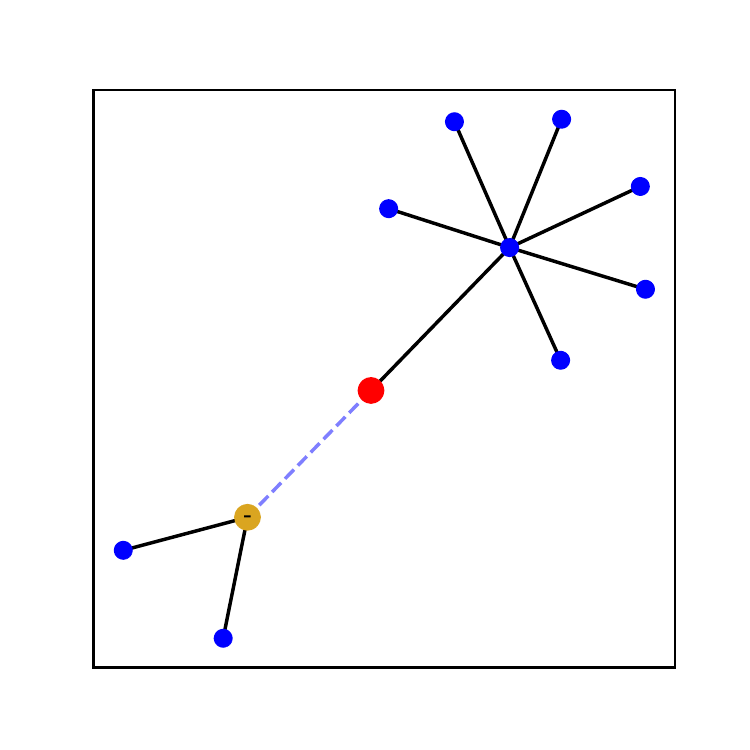} }}

\caption{Case Studies on Localized Graphs.}\label{fig:repre-graph}
\vspace{-0.35in}
\end{figure}

The key to the success of LGCF is that the extracted localized graphs for positive and negative pairs are distinguishable. In other words, they have distinct structures. Thus, we visualize the extracted localized graphs for two positive user-item pairs and their corresponding negative pairs in Figure~\ref{fig:repre-graph}. Note that these two pairs are correctly predicted by LGCF, while wrongly predicted by LightGCN. For the figure, we observe that the localized graphs for positive pairs are substantially different from these of negative pairs, and thus LGCF can distinguish them and make correct predictions. These case studies not only validate the feasibility of using localized graphs for recommendations but also further demonstrate that recommendations via localized graphs are complementary to these via embeddings. 

\section{Related work.}\label{sec:related work}
Collaborative Filtering (CF) is one of the most popular techniques for recommendation~\cite{aggarwal2016model}. Its core idea is to make predictions of a user's preference by analyzing its historical interests. There are basically two types of CF methods: memory-based CF methods and model-based models. Memory-based CF methods aim at making predictions by memorizing its similar users' (or items') historical interactions~\cite{hofmann2004latent}. Model-based CF methods~\cite{koren2009matrix}  aim at predicting by inferring from an underlying model.
With the rapid development of deep neural networks (DNNs), there are more and more works exploring incorporating DNN techniques into model-based CF methods~\cite{he2017neural,xue2017deep}. 
As DNN successfully generalizes to graphs, graph neural networks (GNNs) have drawn great attention. User-item interactions in recommendations can be naturally denoted as a bipartite graph. Thus, more and more works study how to utilize GNN techniques in model-based CF methods including HOP-Rec~\cite{yang2018hop}, PinSage~\cite{ying2018graph}, NGCF~\cite{wang2019neural} and LightGCN~\cite{he2020lightgcn}. Specifically, HOP-Rec~\cite{yang2018hop} includes multi-hop connections into the MF method. PinSage~\cite{ying2018graph} is developed based on GraphSage~\cite{hamilton2017inductive}, and it can be directly applied in industry-level recommendations. NGCF~\cite{wang2019neural} proposes to explicitly encode high-order connectivity into representation via graph filtering operation. LightGCN~\cite{he2020lightgcn} simplifies the design of GNN specially for the recommendation task by eliminating feature transformation and activation function.

\section{Conclusion.}

In this paper, we propose a GNN-based CF model LGCF from a novel perspective for recommendations. It aims at encoding the collaborative filtering information into a localized graph representation. Different from existing embedding-based CF models, LGCF does not require learning embeddings for each user and item. Instead, it focuses on learning the recommendation-related patterns from the localized graph extracted from the input user-item interaction graph for a specific user-item pair. We have conducted extensive experiments to demonstrate the effectiveness of LGCF in recommendation tasks. In the future, we will explore other graph filter and pooling operations to learn the localized graph representations. In addition, we plan to explore some advanced techniques such as attention mechanism to further incorporate different types of GNN-based CF models.

\bibliographystyle{sdm}
\bibliography{sdm}
\end{document}